# Prediction of Stable Ground-State Binary Sodium-Potassium Interalkalis under High Pressures

*Yangmei Chen[1], Xiaozhen Yan*[1,2], Huayun Geng*[2], Xiaowei Sheng[3], Leilei Zhang[2], Hao Wang[2], Jinglong Li[2], Ye Cao,[2] and Xiaolong Pan[2]*

1. School of Science, Jiangxi University of Science and Technology, Ganzhou 341000, Jiangxi, People's Republic of China

2. National Key Laboratory of Shock Wave and Detonation Physics, Institute of Fluid Physics, CAEP, P.O. Box 919-102, Mianyang 621900, Sichuan, People's Republic of China

3. Department of Physics, Anhui Normal University, Anhui 241000, Wuhu, People's Republic of China

**Corresponding Author**

* Xiaozhen Yan: xzyan01@qq.com.

* Huayun Geng: s102genghy@caep.cn.

ABSTRACT The complex structures and electronic properties of alkali metals and their alloys provide a natural laboratory for studying the interelectronic interactions of metals under compression. A recent theoretical study (*J. Phys. Chem. Lett.* **2019**, 10, 3006) predicted an interesting pressure-induced decomposition-recombination behavior of the Na$_2$K compound over a pressure range of 10 - 500 GPa. However, a subsequent experiment (*Phys. Rev. B* **2020**, 101, 224108) reported the formation of NaK rather than Na$_2$K at pressures above 5.9 GPa. To address this discordance, we study the chemical stability of different stoichiometries of Na$_x$K ($x$ = 1/4, 1/3, 1/2, 2/3, 3/4, 4/3, 3/2 and 1 - 4) by effective structure searching method combined with first-principles calculations. Na$_2$K is calculated to be unstable at 5 - 35 GPa due to the decomposition reaction Na$_2$K→ NaK + Na, coinciding well with the experiment. NaK undergoes a combination-decomposition-recombination process accompanied by an opposite charge-transfer behavior between Na and K with pressure. Besides NaK, two hitherto unknown compounds NaK$_3$ and Na$_3$K$_2$ are uncovered. NaK$_3$ is a typical metallic alloy, while Na$_3$K$_2$ is an electride with strong interstitial electron localization.

**KEYWORDS.** Alkali metals; Structure search; First-principles calculation; Crystal structure; Electronic property.

■ INTRODUCTION

The alkali metals with single valence electron configuration are considered textbook examples of electronic structures of metals. At ambient conditions, the quasifree valence electron leads to simple atomic bonding and highly symmetric bcc structures. On modest compression, they retain their free-electron-like behavior and transform to a more close-packed fcc structure. Under further compression, all alkalis undergo phase transitions from fcc to complex structures[1-11] accompanied by a decrease in symmetry, coordination number and packing density.

From the energy-band theory, the pressure induced $s \to p$ and $s \to d$ charge transfer or hybridization is the key to comprehend the appearance of complex phases and properties. [2-3, 12-13]

For the elemental alkali metals, increasing pressure results for Li and Na in $s \to p$ charge transfer, [2, 5, 14-16] and for K, Rb and Cs $s \to d$ transfer. [15, 17] Since the shapes of $p$ and $d$ orbitals are more complex, the $p$- and $d$-governed bonding are inevitably more complex, which therefore results in more complex structures, accompanied by remarkable physical phenomena such as enhanced superconductivity,[18-20] reduced melting temperatures,[4, 21-22] and transformations into electride compounds.[2, 21, 23-26]

For the binary combinations of alkali metals, the underlying charge transfer and redistribution of electrons between them, and the resultant properties have also attracted a lot of attention. An example is the Li-Cs alloy which is not formed at ambient conditions due to the large disparity in their atomic sizes. Theoretical calculations within density-functional theory (DFT) performed by Zhang et al. [27] reveal that compressing can fundamentally alter the repulsive nature of Li and Cs, driving them to form intermetallic compounds LiCs and $Li_7Cs$ at 160 GPa and 80 GPa, respectively. Analysis of valence charge density shows electron transfer from Cs to Li, resulting in oxidation states of Li. Botana et al. [28] predict by structure searching simulations and DFT calculations that Cs can gain electrons from Li, and to form stable compounds $Li_nCs$ ($n$ = 1 - 5) above 100 GPa. However, a subsequent diamond-anvil-cells (DAC) experiment performed by Desgreniers et al. [29] find that Li-Cs alloys could be experimentally synthesized at very low pressure (> 0.1 GPa).

Besides Li-Cs, compound formation has also been reported experimentally[30] in Na-Cs ($Na_2Cs$) and K-Cs ($K_2Cs$, $K_7Cs_6$) systems. The $Na_2Cs$ and $K_2Cs$ compounds in the $MgZn_2$-type Laves phase are calculated to have negative formation enthalpies within DFT. [27] The Li-Na system is predicted theoretically to form an insulating electride compound LiNa under 355 GPa. [31] In the Na-K system, additionally, the eutectic NaK is a liquid alloy at room temperature and solidifies at 260 K, which is used as a coolant in nuclear reactors. [32] When cooling down to 240 K under normal pressure, only the $Na_2K$ solid is observed experimentally. [33] Theoretical calculations [34] confirm the stability of $Na_2K$ in the $MgZn_2$-type Laves phase due to the negative enthalpy. It has also been suggested by limited experimental data on the formation of a $NaK_2$ intermetallic compound at high pressures; however, the existence of this compound has not yet been confirmed. [33]

Most recently, Yang et al. [35] uncovered several new structures of $Na_2K$ over a pressure range of 10 - 500 GPa, by swarm-intelligence structure-searching simulation.

Notably, this compound is predicted to undergoes a decomposition-recombination behavior with pressure. A subsequent DAC experiment [36] observed the formation of $Na_2K$ at pressures below 5.9 GPa. Above this pressure, however, the Na-K system forms NaK instead of $Na_2K$ up to at least 48 GPa, which is inconsistent with the theoretical prediction.

To gain further insight into the chemical stability of Na-K system, we perform extensive structure searches for different stoichiometric $Na_xK$ ($x$ = 1/4, 1/3, 1/2, 2/3, 3/4, 4/3, 3/2 and 1 - 4) under high pressures. Our results reveal that NaK undergoes a combination-decomposition-recombination process with pressure. $Na_2K$ is not stable with respect to NaK and Na. Instead, two ground-state stable phases of $NaK_3$ and $Na_3K_2$ are identified.

■ COMPUTATIONAL DETAILS

Our structural searching simulation is based on a global minimization of *ab initio* total-energy calculations as implemented in the CALYPSO (Crystal structure AnaLYsis by Particle Swarm Optimization) code, [37-41] which has demonstrated good efficiency in predicting high-pressure structures of pure alkali metals, [24, 42] alkali metal polyhydrides [43] and alkali alloys. [28, 31, 35] Structure predictions of $Na_xK$ ($x$ = 1/4, 1/3, 1/2, 2/3, 3/4, 4/3, 3/2 and 1 - 4) are performed at 100, 300 and 500 GPa, with 1 - 4 formula units per simulation cell. Each search generation contains 30-50 structures and the structure searching simulation is usually stopped after generates 900-1500 structures. *Ab initio* structural relaxations and electronic structure calculations are carried out by using the Vienna ab-initio simulation package (VASP) [44] with the

Perdew-Burke-Ernzerhof (PBE) generalized gradient approximation (GGA) functional.[45] $2s^22p^63s^2$ of Na and $3s^23p^64s^1$ of K are treated as valence electrons for projected-augmented-wave pseudopotentials. The cut-off energy for the expansion of wavefunctions into plane waves is set to 950 eV in all calculations, and the Monkhorst-Pack grid with a maximum spacing of 0.02 Å$^{-1}$ is individually adjusted in reciprocal space to the size of each computational cell, which usually give total energies converged to ~1 meV per atom. Lattice dynamics is calculated by the small displacement method as implemented in the PHONOPY package.[46] The formation enthalpy ($\Delta H$) of Na$_m$K$_n$ with respective to elemental Na and K is calculated by using the following formula:

$$\Delta H(\text{Na}_m\text{K}_n) = [H(\text{Na}_m\text{K}_n) - mH(\text{Na}) - nH(\text{K})]/(m+n)$$

wherein $H$ is the enthalpy of the most stable structure of certain compositions at the given pressure. For Na, the bcc, fcc, cI16, oP8, and hP4 structures were adopted[2, 4, 10, 47-48]. And for K, the bcc, fcc, oP8, tI4, oC16, and hP4 structures were adopted[48-50].

■ RESULTS AND DISCUSSION

The structure-searching results of Na$_x$K ($x$ = 1/4, 1/3, 1/2, 2/3, 3/4, 4/3, 3/2 and 1 - 4) at pressures of 100, 300 and 500 GPa are summarized in Figure 1a, wherein the phases located on the convex hull are chemically stable against any type of decomposition. It is apparent that no compounds considered here are stable at 100 GPa because of the positive formation enthalpies. At 300 GPa, NaK$_3$ and Na$_3$K$_2$ become stable, the other compounds NaK, NaK$_2$, NaK$_4$, Na$_2$K$_3$ and Na$_3$K$_4$, as well as the previously predicted Na$_2$K[35] are metastable due to their negative formation enthalpies but locating above the convex hull. When pressure increases to 500 GPa, NaK$_3$ remains its stability, whereas Na$_3$K$_2$ is unstable upon decomposition into NaK and Na.

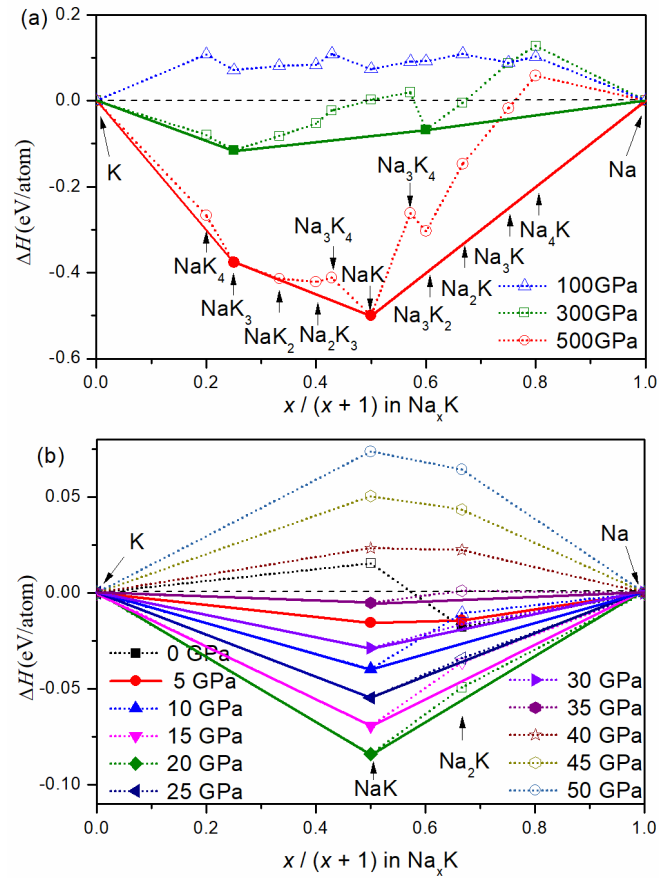

Figure 1. Chemical stabilities of Na-K system with respect to elemental Na and K at different pressures. (a) Chemical stabilities of various Na$_x$K compounds at 100 - 500 GPa (b) Chemically stabilities of NaK and Na$_2$K at 0 - 50 GPa. The stable phases are shown using solid symbols connected by solid lines on the convex hull.

For the simplest stoichiometry of NaK, our searches show that the energetically most favorable structures adopt the $Fd\bar{3}m$ symmetry at 100 and 300 GPa, and the *Ibmm* at 500 GPa. As is shown in Figure 2a, the $Fd\bar{3}m$ structure is comprised of two interpenetrating diamond-type sublattices of Na and K. Both Na and K in this structure have a coordination number of 4. In the *Ibmm* structure (Figure 2b), the coordination numbers of Na and K both increase to 8. The structural details at selected pressures for

$Fd\bar{3}m$ and *Ibmm* phases are listed in Table S1 in the Supporting Information. Phonon calculations reveal no imaginary frequency in the phonon spectrums (Figure S1), implying that these structures are dynamically stable.

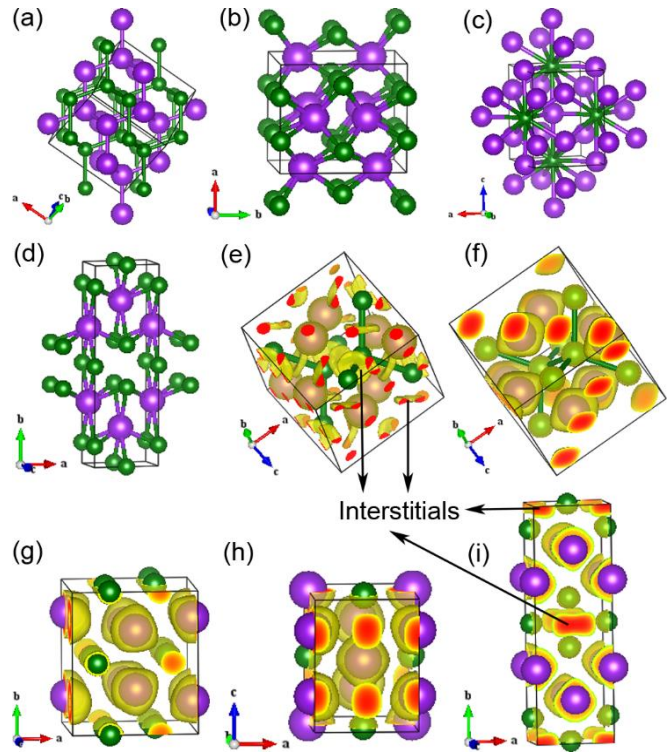

Figure 2. Crystal structures and bonding properties. Crystal structures of (a) $Fd\bar{3}m$ NaK, (b) *Ibmm* NaK, (c) *Pnmm* NaK$_3$ and (d) *Cmmm* Na$_3$K$_2$; Electron localization function plots with isosurface of 0.6 for (e) $Fd\bar{3}m$ NaK at 10 GPa, (f) $Fd\bar{3}m$ NaK at 310 GPa, (g) *Ibmm* NaK at 500 GPa, (h) *Pnmm* NaK$_3$ at 300 GPa and (i) *Cmmm* Na$_3$K$_2$ at 300 GPa. Green (small) and purple (large) spheres denote Na and K atoms, respectively.

As observed in the experiment,[36] the Na-K system forms Na$_2$K at 0 - 5.9 GPa. Above this pressure, however, it forms NaK instead of Na$_2$K up to at least 48 GPa. As revealed by our calculation, NaK becomes stable at a pressure approaches to 5 GPa (Figure 1b). At 35 GPa, it becomes unstable upon decomposition into Na and K. In the case of Na$_2$K, it remains stable at 0 - 5 GPa. Above this pressure, it is unstable due to the decomposition of Na$_2$K → NaK + Na. Therefore, our results correspond well with the experimental observation.[36] Moreover, when pressure further increases to 500 GPa, NaK comes back to the most stable phase on the convex hull (Figure 1a).

It is known that the formation enthalpy of a compound can be separated into two parts of contribution, the relative internal energy $\Delta U$ and the $\Delta PV$ term, i.e., $\Delta H = \Delta U + \Delta PV$. The pressure dependence of $\Delta U$ and $\Delta PV$ of NaK are plotted in Figure 3. Around 5 GPa, $\Delta U$ is calculated to be positive while $\Delta PV$ is negative. The negative $\Delta PV$ suggests that a better packing can be achieved in NaK than in elemental Na and K, which is responsible for its stability. As pressure increases to above 22 GPa, $\Delta PV$ becomes positive and the main contribution of the negative formation enthalpy comes from internal energy $\Delta U$. However, as pressure increases to above 36 GPa (< 300 GPa), the negative $\Delta U$ is not enough to balance the increase in $\Delta PV$, which leads to the decomposition due to the positive formation enthalpy. Furthermore, at pressures above 300 GPa, the formation enthalpy falls back to negative owing to the optimum structure-packing (relative to elemental Na and K) which results in negative $\Delta PV$ (Figure 3b).

In elemental alkali metals, the pressure induced $s \rightarrow p$ and $s \rightarrow d$ charge transfer or hybridization is the key to comprehend the appearance of complex phases and properties.[2-3, 12] To get further insights into the physical origin of the formation of NaK, the nature of chemical bonding is then analyzed.

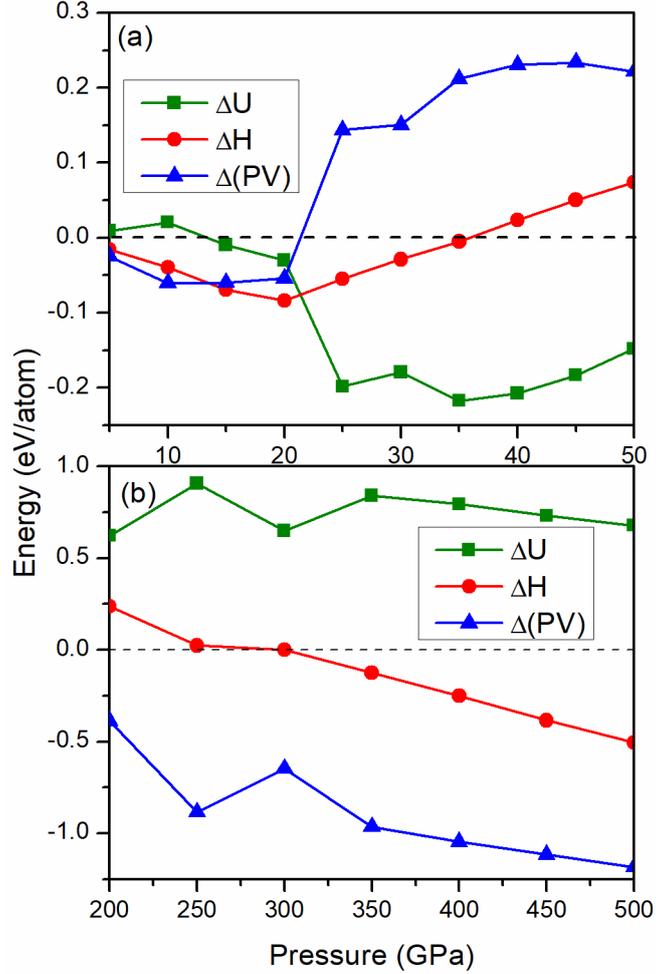

Figure 3. Relative changes in enthalpy H, the PV term, and internal energy U for NaK (a) at 5－50 GPa and (b) 200－500 GPa.

Our Bader charge analysis reveals that the charge-transfer directions between Na and K atoms in NaK are opposite below and above ~50 GPa (Figure S2). The pressure-induced charge reversal has also been found in Na$_2$K[35]. This behavior can be explained by the pressure-induced change in electronegativity of Na and K elements. According to the electronegativity data calculated by Dong et al[51], under low pressures (< 50 GPa), Na element is more electronegative than K since the former has a smaller atomic size, and thus charge transfer from K to Na takes place; Under high pressures (> 50 GPa), K becomes a $d$-block element due to the pressure-induced $s \rightarrow d$ transition, while Na does not, this makes K more electronegative than Na and leads to the inverse charge transfer. It should be noted that Rahm et al[52] have also perform a systematic study on electronegativity of all the elements under pressure, who find that Na is the most electronegative elements in the alkali metals. This is at odds with Dong's results. As described by the authors, this discrepancy may arise from different electronegativity definition, model construction or level of theory. The present Na-K system corresponds to the Dong's model which has also been mentioned in Ref [35].

Furthermore, we also calculate the electron-localization function (ELF) to characterize the atomic bonding. As shown in Figure 2e, at 10 GPa, the ELF basins surround the Na and K atoms correspond to the core electrons. Between the Na and K atoms, the ELF values are less than 0.5, which is typical for ionic and metallic bonding. We also observe ELF maximums (> 0.5)—usually represent interstitial electron localization—situated in the tetrahedral interstitial sites around Na atoms. As pressure increases, the ELF maximums disappear gradually, indicating that the interstitial electron localization is suppressed by pressure (Figure 2f and g). This is in contrast to elemental alkalis, wherein compression would increase interstitial electron localization due to the core-valence overlap [53-56]. This phenomenon has also been found by Frost et al,[36] who explained that the reduced localization with pressure is due to the pressure-induced decrease sizes of interstitial spaces, and they inferred that it may reappear at higher pressures if this system remains stable. However, our calculations show that although NaK is stable under high pressure, no such electron localizations are found neither in the $F d \bar{3} m$ phase at 300 GPa (Figure 2f) nor $Ibmm$ at 500 GPa (Figure 2g).

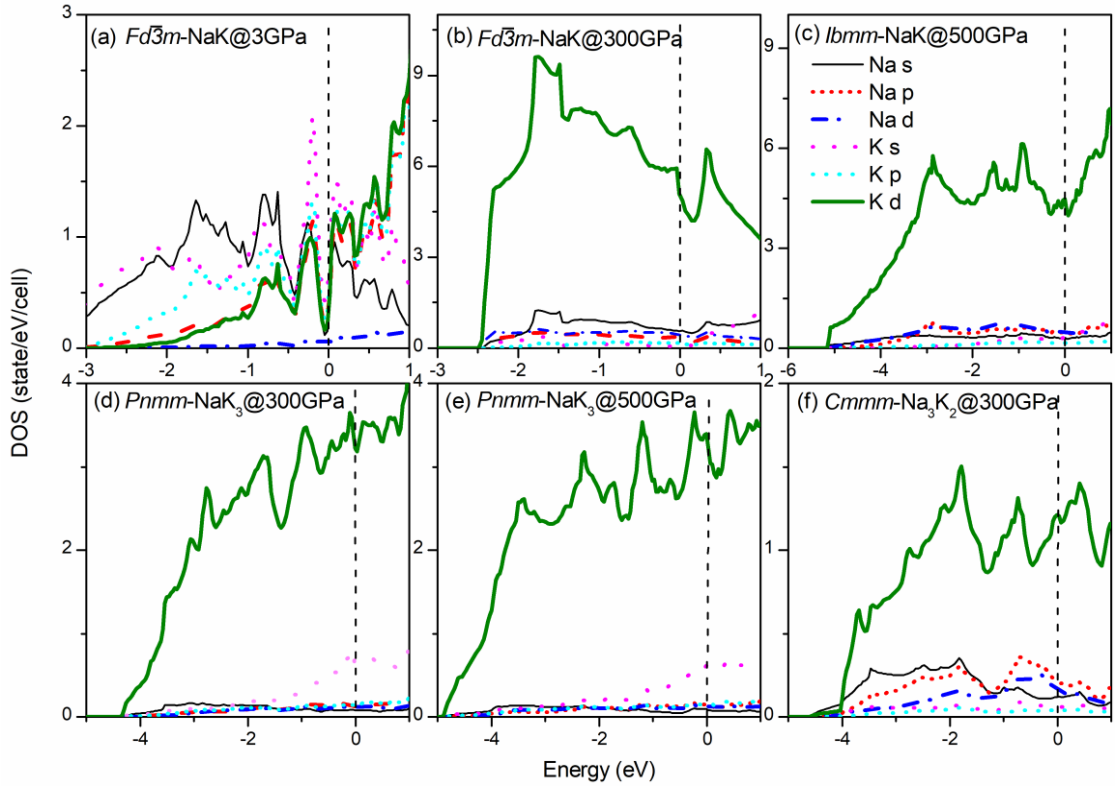

Figure 4. Calculated total and projected density of states of (a) $Fd\bar{3}m$ NaK at 3 GPa, (b) $Fd\bar{3}m$ NaK at 300 GPa, (c) *Ibmm* NaK at 500 GPa, (d, e) *Pnmm* $NaK_3$ at 300 and 500 GPa and (f) *Cmmm* $Na_3K_2$ at 300 GPa.

Figure 4a-c display the calculated electronic density of states (DOS) for NaK at selected pressures. It is apparent that all the DOS results reveal metallic nature of NaK. For $Fd\bar{3}m$ NaK at 3 GPa, the states around the Fermi level are mainly contributed by the *s* and *p* orbitals of Na and K, and also have significant K-*d* contribution. As pressure increases, the K-*d* contribution becomes dominated (Figure 4b). This is expected since the pressure-induced $s \rightarrow d$ transition is common in K element [17, 52]. At 500 GPa, the DOS of *Ibmm* phase is analogously, wherein the K-*d* orbital contributes dominantly to the states around the Fermi level. It is interesting to note that the DOS of $Fd\bar{3}m$ NaK at low pressures is characterized by a pronounced pseudogap in the vicinity of Fermi level (Figure 4a). This pseudogap does not present at high pressures (see Figure 4b and c).

The appearance of pseudogap may be attributed to the localized electrons in the interstitial.

NaK$_3$ is predicted to crystallize in a Cu$_3$Ge-type structure (space group *Pnmm*) with 2 formula units per unit cell, consisting of K-sharing 12-fold NaK$_{12}$ icosahedron (Figure 2c). As is evidenced by the large DOS at the Fermi level (Figure 4d and e), this compound is a typical metallic alloy. Both at 300 and 500 GPa, the states around the Fermi level show large components of K-*d* contribution. The distances of neighboring Na-Na, Na-K and K-K distances are 3.19, 2.15 and 2.18 Å at 300 GPa, respectively. Given that the atomic radii of Na and K are, respectively, 1.18 and 1.23 Å, at 300 GPa,[57] the core-core overlaps in neighboring Na-K and K-K would take place. As a results, the dispersions of semi-core bands of K 3*s* and 3*p* are increased obviously (see Figure S3). It is well known that such core-core overlap is common in materials under high pressures, which plays an important role in their stability[2, 21, 56, 58-59]. The ELF results (Figure 2h) indicate that the main interactions in this system are ionic and metallic.

In the case of Na$_3$K$_2$, it stabilizes in a *Cmmm* structure (Figure 2d) at 300 GPa and decomposes into NaK and K at 500 GPa. In *Cmmm* Na$_3$K$_2$, both Na and K are 8-coordinated. Each K is coordinated to 8 Na atoms and each Na to 4 K and 4 Na atoms. Na$_3$K$_2$ is also metallic with large value of DOS at the Fermi level contributed dominantly from the K-*d* states (Figure 4f). Comparing the atomic distances (the nearest neighboring Na-Na, Na-K and K-K distances are 1.96, 2.13 and 2.09 Å, respectively) with sums of atomic radii of Na and K at 300 GPa indicates that the core-core overlap also occurs in this system. In contrast to NaK$_3$, the ELF plot of Na$_3$K$_2$ indicates strong interstitial electron localization (Figure 2i). Bader charge analysis also confirms this conclusion, which reveals that the charges of Na, K atoms and the interstitial sites are approximatively 0.6, -0.5 and -0.9 respectively at 300 GPa, indicating that the electrons delocalized at the interstitial sites are mainly transferred from Na. These results demonstrate that $Na_3K_2$ is an electride compound in term of $Na_3K_2(e)$.

■ CONCLUSIONS

In summary, the swarm-intelligence structure-searching method is employed to explore the stability of Na-K system under high pressure. Three stoichiometries of NaK, $NaK_3$ and $Na_3K_2$ are found to form ground-state compounds under different pressures. NaK within the $Fd\bar{3}m$ structure is identified to stabilize at 5 - 35 GPa, and becomes unstable upon decomposition into Na and K above this pressure. When pressure further increase to 500 GPa, NaK restabilizes in a $Ibmm$ phase. The most K rich compound $NaK_3$ stabilizes in a $Cu_3Ge$-type structure at 300 - 500 GPa. $Na_3K_2$ is predicted to crystallize in a $Cmmm$ structure at 300 GPa and decomposes into NaK and K at 500 GPa. Remarkable pressure-induced charge-transfer reversal between Na and K atoms are revealed in these systems. Our results will deepen the understanding of interelectronic interactions of metals and stimulate future experimental and theoretical investigations.

■ ASSOCIATED CONTENT

**Supporting Information**.

Structure details, Phonon spectrums, Bader charge, Projected DOS, Pressure dependent enthalpies of formation of NaK, $NaK_3$ and $Na_3K_2$.

## ■ AUTHOR INFORMATION

**Corresponding Authors**

* E-mail: xzyan01@qq.com.

* E-mail: s102genghy@caep.ac.cn.

**Notes**

The authors declare no competing financial interests.

## ■ ACKNOWLEDGMENT

This work is supported by the National Natural Science Foundation of China under Grant nos. 11704163, 11672274, 11804131, and 11274281, the CAEP Research Projects under Grant nos. 2012A0101001 and 2015B0101005, the NSAF under Grant no. U1730248, the Fund of National Key Laboratory of Shock Wave and Detonation Physics of China under Grant no. 6142A03010101, the China Postdoctoral Science Foundation under Grant no. 2017M623064, the Natural Science Foundation of Jiangxi Province of China under Grant no. 20181BAB211007 and the PhD research startup foundation of Jiangxi University of Science and Technology under Grant nos. 3401223301, 3401223256.  This work is supported by the National Natural Science Foundation of China (11804131, 11704163, 11672274 and 11404006), the Science challenge Project (TZ2016001), the CAEP Research Project (CX2019002), the NSAF (U1730248), the Science challenge Project (TZ2016001), the CAEP Research Project (CX2019002), the China Postdoctoral Science Foundation (2017M623064), the Natural Science Foundation of Jiangxi Province of China (20181BAB211007), and the Ph.D. research startup foundation of Jiangxi University of Science and Technology

(3401223301 and 3401223256), the Scientific Research Fund of Jiangxi Provincial Education Department (Grant No.GJJ160594).

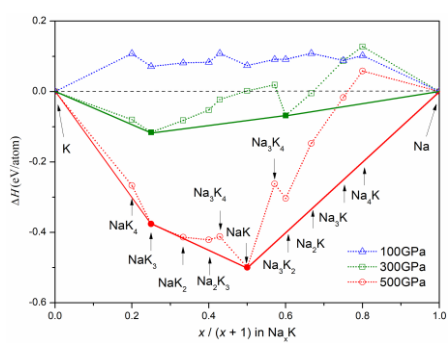

Table of contents (TOC) graphic: Chemical stabilities of Na-K system with respect to elemental Na and K at different pressures. The compounds of NaK, $NaK_3$ and $Na_3K_2$ are predicted to be stable under different pressures.

# Supporting Information for "Prediction of Stable Ground-State Binary Sodium-Potassium Interalkalis under High Pressures"

Yangmei Chen[1], Xiaozhen Yan*[1,2], Huayun Geng*[2], Xiaowei Sheng[3], Leilei Zhang[2], Hao Wang[2], Jinglong Li[2], Ye Cao,[2] and Xiaolong Pan[2]

1. School of Science, Jiangxi University of Science and Technology, Ganzhou 341000, Jiangxi, People's Republic of China

2. National Key Laboratory of Shock Wave and Detonation Physics, Institute of Fluid Physics, CAEP, P.O. Box 919-102, Mianyang 621900, Sichuan, People's Republic of China

3. Department of Physics, Anhui Normal University, Anhui 241000, Wuhu, People's Republic of China

**Corresponding Author**

* Xiaozhen Yan: xzyan01@qq.com.

* Huayun Geng: s102genghy@caep.cn.

**Auxiliary Table S1**. Structure details of the conventional unit cell of NaK, NaK$_3$ and Na$_3$K$_2$ from the CALYPSO structure searches at different pressures.

| Phase | Pressure (GPa) | Space group | Lattice parameters | Atomic coordinates |
|---|---|---|---|---|
| NaK | 10 | $Fd\bar{3}m$ | a = b = c = 6.7242Å | Na(8a) 0.75 -0.25 0.25 |
| | | | α = β = γ = 90.000º | K(8b) 0.25 -0.25 0.25 |
| NaK | 500 | *Ibmm* | a = 5.0839 Å | Na(8h) -0.50 0.03 -0.70 |
| | | | b = 4.2379 Å | K(8i) -0.71 0.75 -0.98 |
| | | | c = 4.4169 Å | |
| | | | α = β = γ = 90.000º | |
| NaK$_3$ | 300 | *Pnmm* | a = 3.4771 Å | Na(2b) 0.50 0.00 -0.32 |
| | | | b = 4.4920 Å | K1(4e) 0.50 0.75 -0.83 |
| | | | c = 3.9227 Å | K2(2a) 0.50 0.50 -0.35 |
| | | | α = β = γ = 90.000º | |
| Na$_3$K$_2$ | 300 | *Cmmm* | a = 3.2827 Å | Na1(4j) 0.00 0.89 0.50 |
| | | | b = 10.2332 Å | Na3(2c) 0.50 0.00 0.50 |
| | | | c = 2.1409 Å | K1(4i) 0.50 0.81 0.00 |
| | | | α = β = γ = 90.000º | |

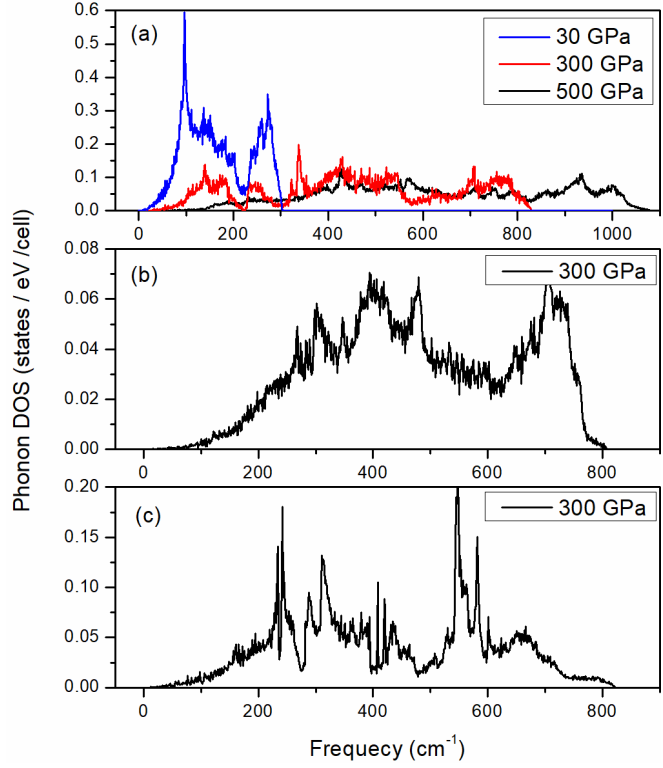

**Figure S1.** Phonon spectrums of (a) $Fd\bar{3}m$ NaK at 30 GPa, 300 GPa and *Ibmm* NaK at 500 GPa, (b) *Pnmm* NaK$_3$ at 300 GPa, (c) *Cmmm* Na$_3$K$_2$ at 300 GPa.

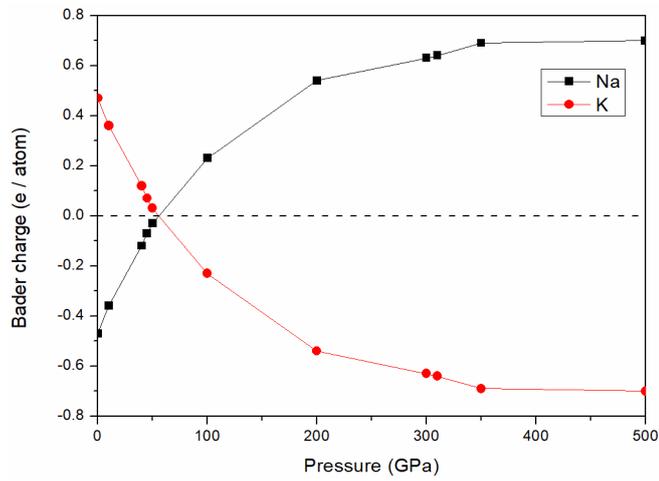

**Figure S2.** Bader charge of Na and K in NaK. Positive/negative value corresponds to loss/gain of electron.

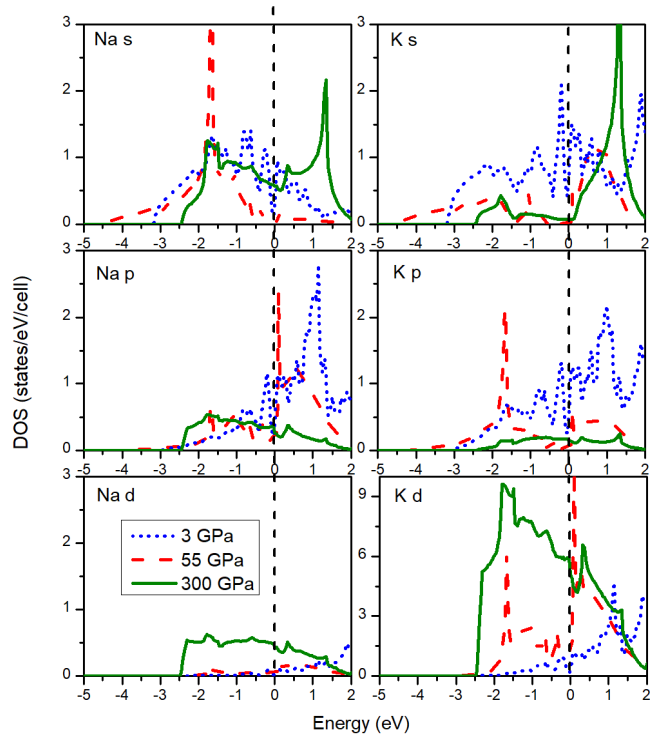

**Figure S3.** Pressure dependence of projected electronic DOS $Fd\bar{3}m$ NaK. The zero of energy is placed at the Fermi energy.

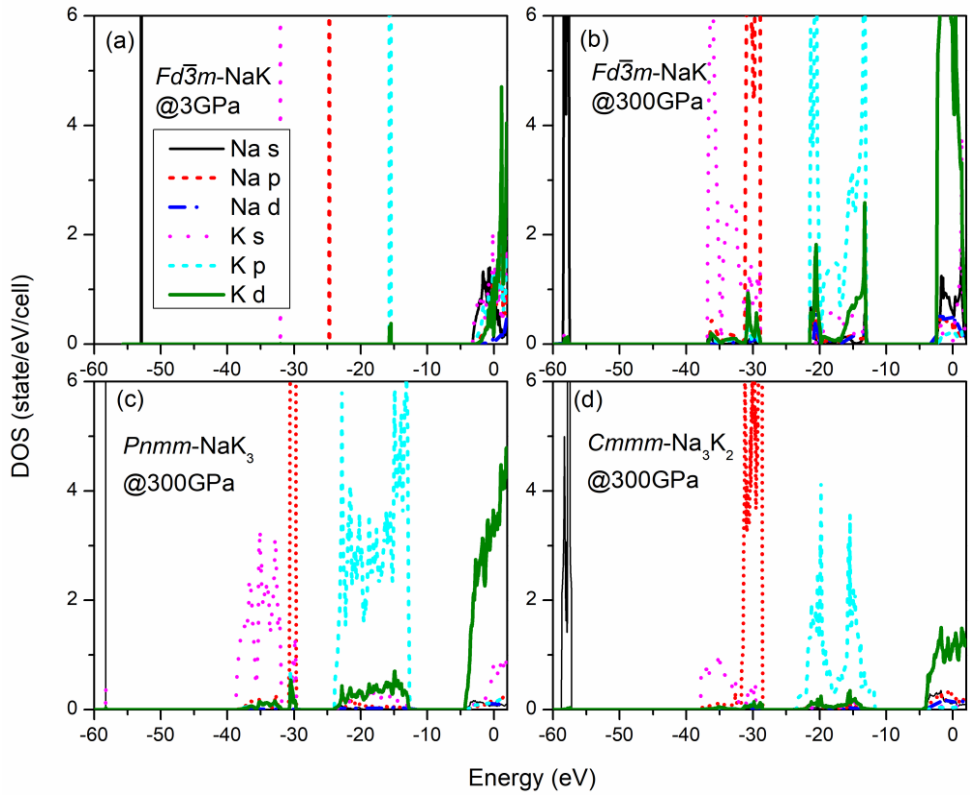

**Figure S4.** Projected electronic DOS showing the Na 2$s$, Na 2$p$, K 3$s$ and K 3$p$

semi-core states and the valence states. The zero of energy is placed at the Fermi energy.

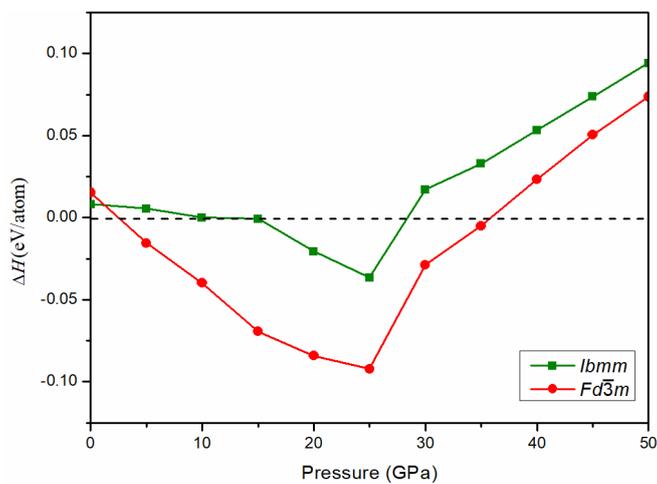

**Figure S5.** Enthalpies of formation of NaK with respect to decomposition into Na and K at 0 - 50 GPa.

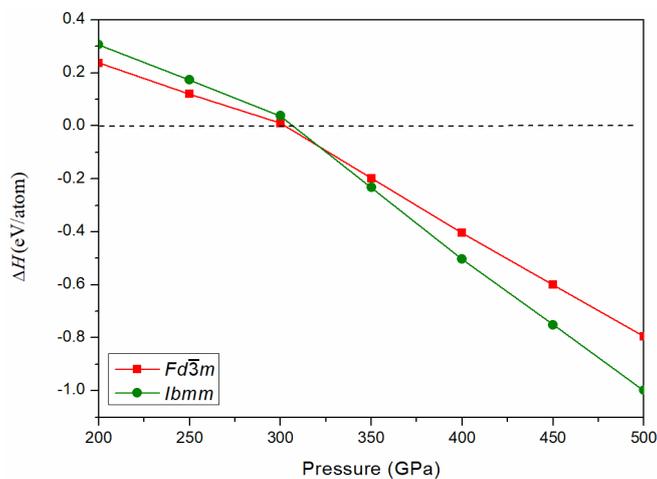

**Figure S6.** Enthalpies of formation of NaK with respect to decomposition into Na and K at 200 - 500 GPa.

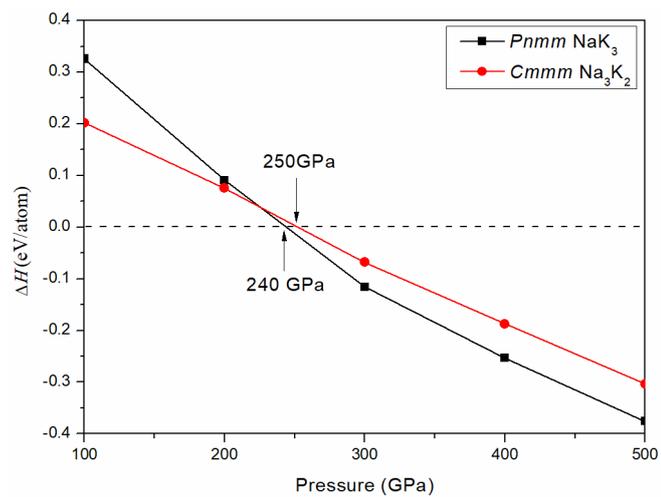

**Figure S7.** Enthalpies of formation of NaK$_3$ and Na$_3$K$_2$ with respect to decomposition nto Na and K at 100 - 500 GPa.